\documentclass[fleqn,usenatbib]{mnras}

\usepackage{newtxtext,newtxmath}
\usepackage[T1]{fontenc}
\usepackage{ae,aecompl}

\usepackage{graphicx}	% Including figure files
\usepackage{amsmath}	% Advanced maths commands
\usepackage{amssymb}	% Extra maths symbols
\usepackage[x11names]{xcolor}

\hypersetup{pdftitle={Neutrino Fluence from Gamma-Ray Bursts: Off-Axis View of Structured Jets},
pdfsubject={Neutrino Fluence from Gamma-Ray Bursts: Off-Axis View of Structured Jets},
pdfauthor={Markus Ahlers and Lea Halser},
pdfstartview={FitH},
colorlinks=true,
bookmarksopen=false,
bookmarksnumbered=false,
bookmarksopenlevel=0}

\usepackage{geometry}
\title[Neutrinos from GRBs: Off-Axis View of Structured Jets]{Neutrino Fluence from Gamma-Ray Bursts:\\Off-Axis View of Structured Jets}

\author[Markus Ahlers \& Lea Halser]{
Markus Ahlers \& Lea Halser
\\
Niels Bohr International Academy \& Discovery Center, Niels Bohr Institute, University of Copenhagen, DK-2100 Copenhagen, DK
}

\date{}

\pubyear{2019}

\begin{document}

\setlength{\mathindent}{1em}
\setlength{\parindent}{1em}

\label{firstpage}
\pagerange{\pageref{firstpage}--\pageref{lastpage}}
\maketitle

\begin{abstract}
We investigate the expected high-energy neutrino fluence from internal shocks produced in the relativistic outflow of gamma-ray bursts. Previous model predictions have primarily focussed on on-axis observations of uniform jets. Here we present a generalization to account for arbitrary viewing angles and jet structures. Based on this formalism, we provide an improved scaling relation that expresses off-axis neutrino fluences in terms of on-axis model predictions. We also find that the neutrino fluence from structured jets can exhibit a strong angular dependence relative to that of $\gamma$-rays and can be far more extended. We examine this behavior in detail for the recent short gamma-ray burst GRB 170817A observed in coincidence with the gravitational wave event GW170817.
\end{abstract}


%%%%%%%%%%%%%%%%%%%%%%%%%%%%%%%%%%%%%%%%%%%%%%%%%%

%%%%%%%%%%%%%%%%% BODY OF PAPER %%%%%%%%%%%%%%%%%%

\section{Introduction}\label{ch1}

Gamma-Ray Bursts (GRBs) are some of the most energetic transient phenomena in our Universe that dominate the $\gamma$-ray sky over their brief existence. The burst duration, ranging from milliseconds to a few minutes, requires central engines that release their energy explosively into a compact volume of space. The GRB data show a bimodal distribution of long ($\gtrsim2$s) and short bursts, indicating different progenitor systems. The origin of long-duration GRBs has now been established as the core-collapse of massive stars~\citep{Woosley:1993wj} by the association with type Ibc supernovae in a few cases~\citep{Hjorth:2011zx}. The recent observation of GRB 170817A in association with the gravitational wave GW170817~\citep{GBM:2017lvd,Monitor:2017mdv} has confirmed the idea that (at least some) short-duration GRB originate from binary neutron star mergers~\citep{1986ApJ...308L..43P,Eichler:1989ve,Narayan:1992iy}. The subsequent multi-wavelength observations of this system also provided evidence for an associated kilonova/macronova from merger ejecta~\citep{Villar:2017wcc} and allowed for a detailed study of the jet structure by the late-time GRB afterglow~\citep{Lazzati:2017zsj,Troja:2018ruz,Margutti:2018xqd,Lamb:2018qfn,Ghirlanda:2018uyx,Lyman:2018qjg}.

After core-collapse or merger, the nascent compact remnant -- a black hole or rapidly spinning neutron star -- is initially girded by a thick gas torus from which it starts to accrete matter at a rate of up to a few solar masses per second. The system is expected to launch axisymmetric outflows via the deposition of energy and/or momentum above the poles of the compact remnant. The underlying mechanism is uncertain and could be related to neutrino pair annihilation powered by neutrino emission of a hyper-accreting disk~\citep{Popham:1998ab,Liu:2017kga} or magnetohydrodynamical processes that extract the rotational energy of a remnant black hole~\citep{Blandford:1977ds}. The interaction of the expanding and accelerating outflow with the accretion torus collimates the outflow into a jet. Subsequent interactions with dynamical merger ejecta or the stellar envelope further collimate and shape the jet until it emerges (or not) from the progenitor environment. We refer to \cite{Zhang:2018ond} for a recent detailed review of the status of GRB observations and models.

For the remainder of this paper, we will assume that the prompt $\gamma$-ray display is related to energy dissipation in the jet via internal shocks~\citep{Rees:1994nw,Paczynski:1994uv}. The variability of the central engine can result in variations of the Lorentz factor in individual sub-shells of the outflow that eventually collide~\citep{Shemi:1990rv,Rees:1992ek,Meszaros:1993tv}. Electrons accelerated by first order Fermi acceleration in the internal shock environment radiate via synchrotron emission, which can contribute to or event dominate the observed prompt $\gamma$-ray display~\citep{Rees:1994nw,Paczynski:1994uv}. In order to be visible, these internal shocks have to occur above the photosphere, where the jet becomes optically thin to Thomson scattering. However, it has been argued that the dissipation of bulk jet motion via (combinations of) internal shocks, magnetic reconnection or neutron-proton collisions close to the photosphere can also produce the typical GRB phenomenology~\citep{Rees:2004gt,Ioka:2007qk,Beloborodov:2009be,Lazzati:2010af}. Eventually, the collision of the fireball with interstellar gas forms external shocks that can explain the GRB afterglow ranging from radio to X-ray frequencies~\citep{Meszaros:1993ft,Meszaros:1994sd}.

Baryons entrained in the jet are inevitably accelerated along with the electrons in internal shocks. \cite{Waxman:1995vg} argued that a typical GRB environment can satisfy the requirements to accelerate cosmic rays to the extreme energies of beyond $10^{20}$~eV observed on Earth. A smoking-gun test of this scenario is the production of high-energy neutrinos from the decay of charged pions and kaons produced by CR interactions with the internal photon background~\citep{Waxman:1997ti,Guetta:2003wi,Murase:2005hy,Zhang:2012qy}. Searches of neutrino emission of GRBs with the IceCube neutrino observatory at the South Pole has put meaningful constraints on the neutrino emission of GRBs~\citep{Ahlers:2011jj,Abbasi:2012zw,Aartsen:2017wea} and has triggered various model revisions~\citep{Murase:2006mm,Li:2011ah,Hummer:2011ms,He:2012tq,Murase:2013ffa,Senno:2015tsn,Denton:2017jwk}. 

Most GRB neutrino predictions are based on on-axis observations of a uniform jet with constant bulk Lorentz factor $\Gamma$ within a half-opening angle $\Delta\theta$ that is significantly larger than the kinematic angle $1/\Gamma$. The apparent brightness of the source is then significantly enhanced due to the strong Doppler boost of the emission. However, the recent observations of GRB 170817A \& GW170817~\citep{GBM:2017lvd,Monitor:2017mdv} and the multi-wavelength emission of its late-time afterglow~\citep{Lazzati:2017zsj} has confirmed earlier speculations that the GRB jet is structured. This explains the brightness of the GRB despite our large viewing angle of $\gtrsim15^\circ$. 

In this paper, we study the neutrino fluence in the internal shock model of GRBs for arbitrary viewing angles and jet structures. In section~\ref{sec1} we will provide a detailed derivation of the relation between the internal emissivity of the GRB and the fluence for an observer at arbitrary relative viewing angles. Our formalism will clarify some misconceptions that have appeared in the literature and provide an improved scaling relation of the particle fluence. In section \ref{sec2} we will study off-axis emission for various jet structures and determine a revised scaling relation that allows to express off-axis fluence predictions based on on-axis models. We then study neutrino emission from internal shocks in structured jets in section \ref{sec3} and show that the emissivity of neutrinos is expected to have a strong angular dependence relative to the $\gamma$-ray display. We illustrate this behavior in section \ref{sec4} for a structured jet model inferred from the afterglow of GRB 170817A before we conclude in section \ref{sec5}. 

Throughout this paper we work with natural Heaviside-Lorentz units with $\hbar=c=\varepsilon_0=\mu_0=1$, $\alpha = e^2/(4\pi)\simeq 1/137$ and $1~{\rm G}\simeq 1.95\times10^{-2}\,{\rm eV}^2$. Boldface quantities indicate vectors.

\section{Prompt Emission from Relativistic Shells}\label{sec1}

The general relation of the energy fluence $\mathcal{F}$ (units of ${\rm GeV}$ ${\rm cm}^{-2}$) from structured jets observed under arbitrary viewing angles can be determined via the specific emissivity\footnote{We changed the standard notation ``$j_\nu$'' for the specific emissivity to avoid confusion with neutrino-related quantities.} $j$ (units of ${\rm cm}^{-3}$ ${\rm s}^{-1}$ ${\rm sr}^{-1}$). This ansatz has been used by \cite{Granot:1998ep}, \cite{1999ApJ...523..187W}, \cite{Nakamura:2001kd} and \cite{Salafia:2016wru} to derive the time-dependent electromagnetic emission of GRBs. The dependence of the isotropic-equivalent energy on jet structure and viewing angle has been studied by \cite{Yamazaki:2003af}, \cite{Eichler:2004ev} and \cite{Salafia:2015vla}. We present here a simple and concise derivation of this relation for thin relativistic shells, also accounting for cosmological redshift. The resulting expressions will allow us to relate the photon density in the structured jet to the observed prompt $\gamma$-ray fluence and to determine the efficiency of neutrino emission from cosmic ray interactions in colliding sub-shells.

\begin{figure}\centering
\includegraphics[width=\linewidth]{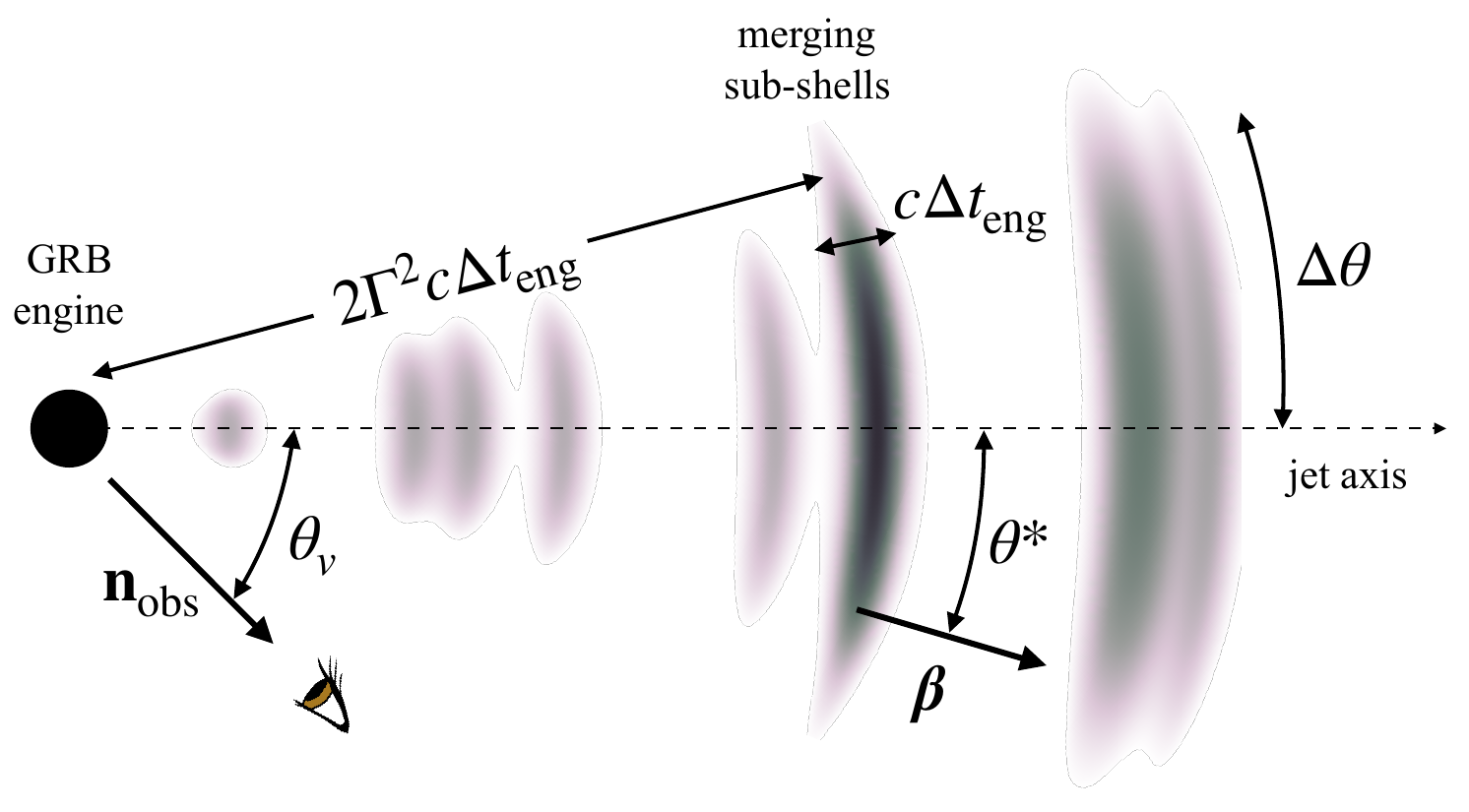}\\[-0.1cm]
\caption[]{Sketch of colliding sub-shells of a variable GRB outflow. Sub-shells with different bulk Lorentz factors $\Delta\Gamma\sim\Gamma$ that are emitted from the GRB engine with time difference $\Delta t_{\rm eng}$ merge at a distance $r_{\rm dis}\sim 2\Gamma^2c\Delta t_{\rm eng}$ and dissipate bulk kinetic energy. Internal shocks accelerate electrons and protons and contribute to the non-thermal emission of the merged shell. Emission along the shell is boosted into the observer frame along a radial velocity vector $\boldsymbol{\beta}(\Omega^*)$. The observer sees the emission under a viewing angle $\theta_v$ in the direction ${\bf n}_{\rm obs}$.}\label{fig1}
\end{figure}

A sketch of the variable GRB outflow and the resulting collision of sub-shells is shown in Fig.~\ref{fig1}. In the observer's rest frame, the fluence per area ${\rm d}A$ from individual elements of a merged shell is related to the emission into a solid angle ${\rm d}\Omega = {\rm d}A/d_A^2$ for a source at angular diameter distance $d_A$. The combined emission of one shell is therefore
\begin{equation}\label{eq:fluenceOBS}
\mathcal{F} = \frac{1}{d_A^2}\int {\rm d}V \int{\rm d} \epsilon\int{\rm d} t j\,.
\end{equation}
The specific emissivity $j$ in the observer's reference frame is related to the specific emissivity $j'$ in the rest frame of the sub-shell (denoted by primed quantities in the following) as~\citep{1979rpa..book.....R}
\begin{equation}\label{eq:j_trafo}
j = \frac{\mathcal{D}^2}{(1+z)^2}j'\,,
\end{equation}
where $z$ denotes the redshift of the source and $\mathcal{D}$ the Doppler factor of the specific volume element. In the following, we will assume that the jet structure in the GRB's rest frame (denoted by starred quantities in the following) is axisymmetric. The spherical coordinate system is parametrized by zenith angle $\theta^*$ and azimuth angle $\phi^*$ such that the jet axis aligns with the $\theta^*=0$ direction. Note that we do not account for the counter-jet in our calculation, but this addition is trivial. At a sufficiently large distance from the central engine, the jet flow is assumed to be radial. The relative viewing angle between the observer and jet axis is denoted as $\theta_{v}$.
The Doppler factor can then be expressed as
\begin{equation}\label{eq:DopplerStar}
\mathcal{D}(\Omega^*) = \left[\Gamma(\theta^*)(1-\boldsymbol{\beta}(\Omega^*)\!\cdot\!{\bf n}_{\rm obs})\right]^{-1}\,,
\end{equation}
where $\boldsymbol{\beta}(\Omega^*)$ corresponds to the radial velocity vector of the specific volume element in the GRB's rest frame and ${\bf n}_{\rm obs}$ is a unit vector pointing towards the location of the observer.  Due to the symmetry of the jet we can express the scalar product in (\ref{eq:DopplerStar}) as
\begin{equation}
\boldsymbol{\beta}(\Omega^*)\!\cdot\!{\bf n}_{\rm obs} = \beta(\theta^*)\left(\sin\theta^*\cos\phi^*\sin\theta_v + \cos\theta^*\cos\theta_v\right)\,.
\end{equation}
Using the transformation of energy $\epsilon' = (1+z)\epsilon/\mathcal{D}$, volume $V' = (1+z)V/\mathcal{D}$ and time $t' = t\mathcal{D}/(1+z)$, we can express Eq.~(\ref{eq:fluenceOBS}) as
\begin{equation}\label{eq:fluenceSHELL} 
\mathcal{F} = \frac{1+z}{d_L^2}\int {\rm d}V' \int{\rm d} \epsilon'\int{\rm d} t'\mathcal{D}^3(\Omega^*) j'\,.
\end{equation}
In this expression we have used the relation $d_L(z) = (1+z)^2d_A(z)$ between the luminosity and angular diameter distance for a source at redshift $z$. 

%%%%%%%%%%%%%%%%%%%%%%%%
%%%%%%%%%%%%%%%%%%%%%%%%

The infinitesimal volume element ${\rm d}V'$ in the rest frame of the sub-shell is related to the volume element ${\rm d}V^*$ in the frame of the central engine as ${\rm d}V' = \Gamma(\theta^*) {\rm d}V^*$. In the internal shock model, the shell radius and width (in the central engine frame) can be related to the variability time scale $\Delta t_{\rm eng}$ of the central engine as $r_{\rm dis}\simeq 2\Gamma^2c\Delta t_{\rm eng}$ and $\Delta r \simeq c\Delta t_{\rm eng}$. The time-integrated emissivity can then be expressed as a sum over $N_{\rm sh}$ merging sub-shells with width $\Delta r$ that appear at a characteristic distance $r_{\rm dis}$,
\begin{equation}
j^*(\theta^*) \simeq N_{\rm sh}\Delta r(\theta^*)\delta(r^*-r_{\rm dis}(\theta^*))j^*_{\rm IC}(\theta^*)\,.
\end{equation}
The total number of colliding sub-shells can be estimated by the total engine activity $T_{\rm GRB}$ as $N_{\rm sh}\simeq \xi T_{\rm GRB}/\Delta t_{\rm eng}$ where we have introduced an intermittency factor $\xi\leq1$. For simplicity, we will assume in the following that the total engine activity is related to the observation time as $T_{\rm GRB}\simeq T_{90}/(1+z)$ and $\xi=1$. Note that the observed variability time-scale $t_{\rm var}$ of a thin jet with viewing angle $\theta_{\rm obs}$ can be related to the engine time scale as $t_{\rm var}/\Delta t_{\rm eng} \simeq \mathcal{D}(0)/\mathcal{D}(\theta_{\rm obs})$, whereas the total observed emission $T_{90}$ is only marginally effected by the off-axis emission~\citep{Salafia:2016wru}.

The specific emissivity $j'_{\rm IC}$ in the rest frame of the sub-shell is assumed to be isotropic. The time-integrated emission can therefore be expressed in terms of a spectral density $n'$:
\begin{equation}
\epsilon'n'(\theta^*) = 4\pi\int{\rm d} t'j_{\rm IC}'(\theta^*)\,.
\end{equation}
The background of relativistic particles in the shell rest frame contributes to the total internal energy density of the shell as
\begin{equation}
u'(\theta^*)  = \int{\rm d}\epsilon'\epsilon' n'(\theta^*)\,. 
\end{equation}
This allows us to express the observed fluence by the internal energy density as: 
\begin{equation}\label{eq:fluencefinal}
\mathcal{F} \simeq \frac{cT_{90}}{4\pi d_L^2}\int {\rm d}\Omega^*\Gamma(\theta^*) \mathcal{D}^3(\Omega^*) r_{\rm dis}^2(\theta^*)u'(\theta^*)\,.
\end{equation}
This is the most general expression for the prompt fluence emitted from a thin shell of an axisymmetric radial jet.

%%%%%%%%%%%%%%%%%%%%%%%%%%%%
\begin{figure*}\centering
\includegraphics[width=0.5\linewidth]{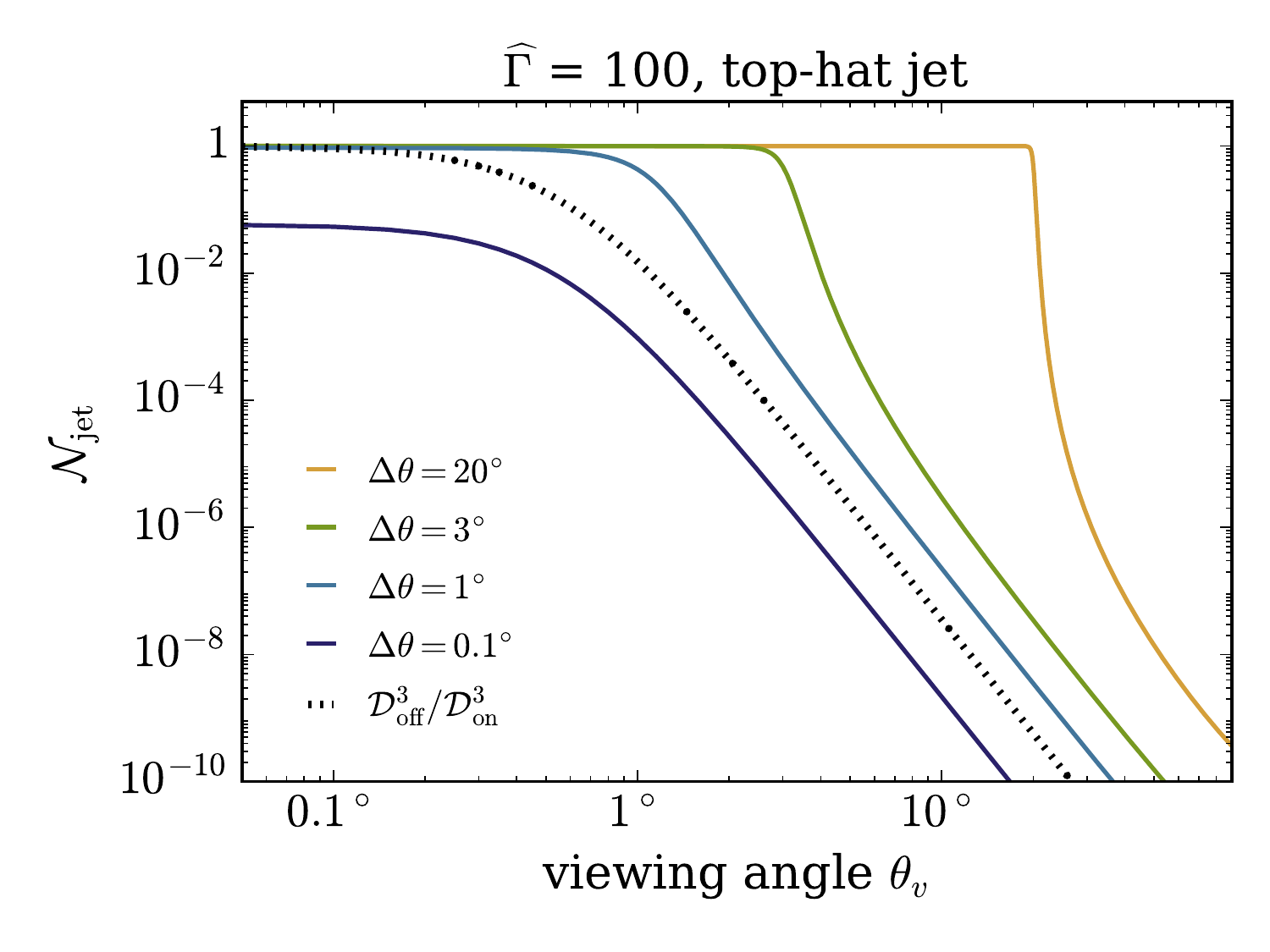}\hfill\includegraphics[width=0.5\linewidth]{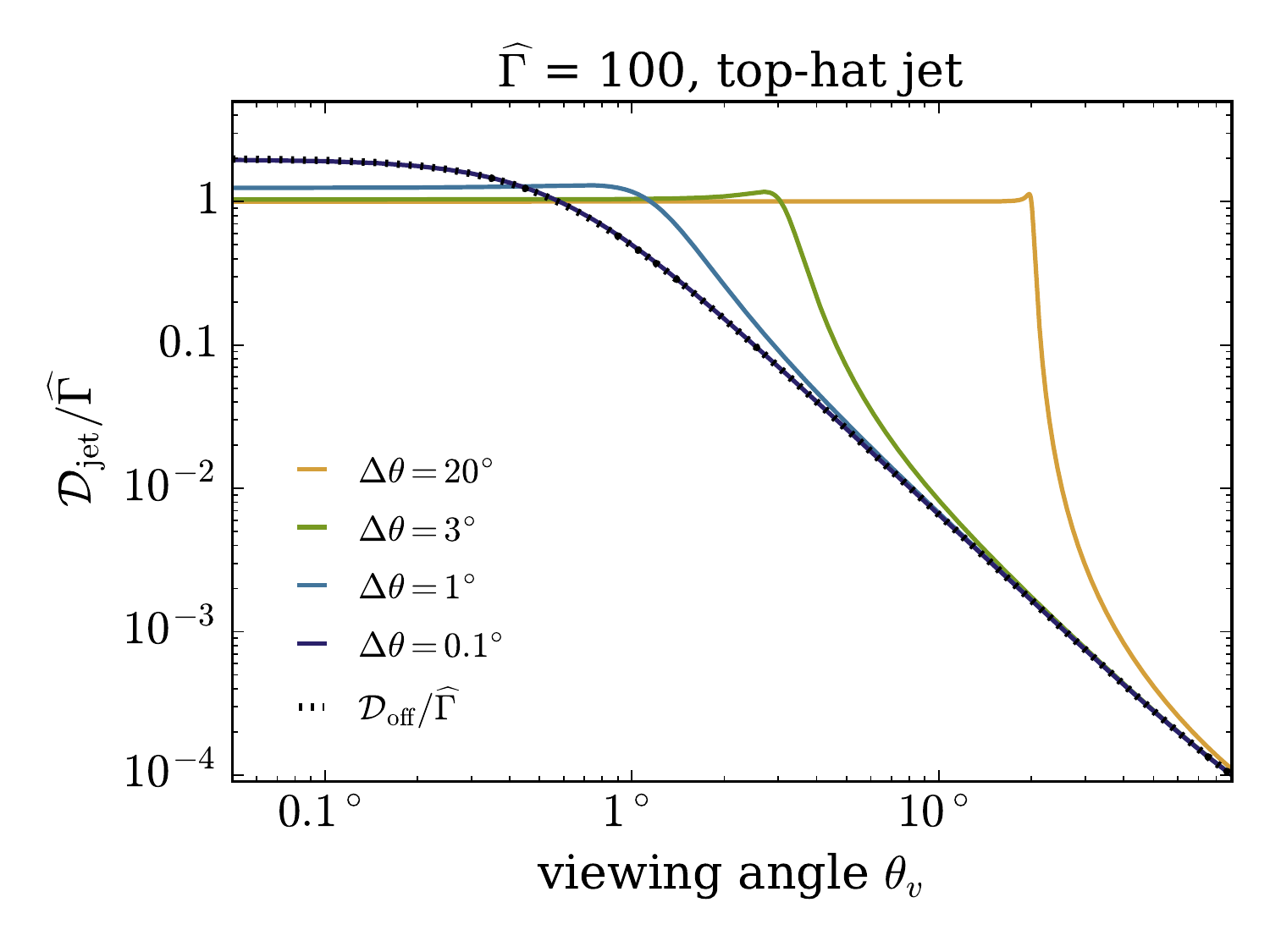}\\
\includegraphics[width=0.5\linewidth]{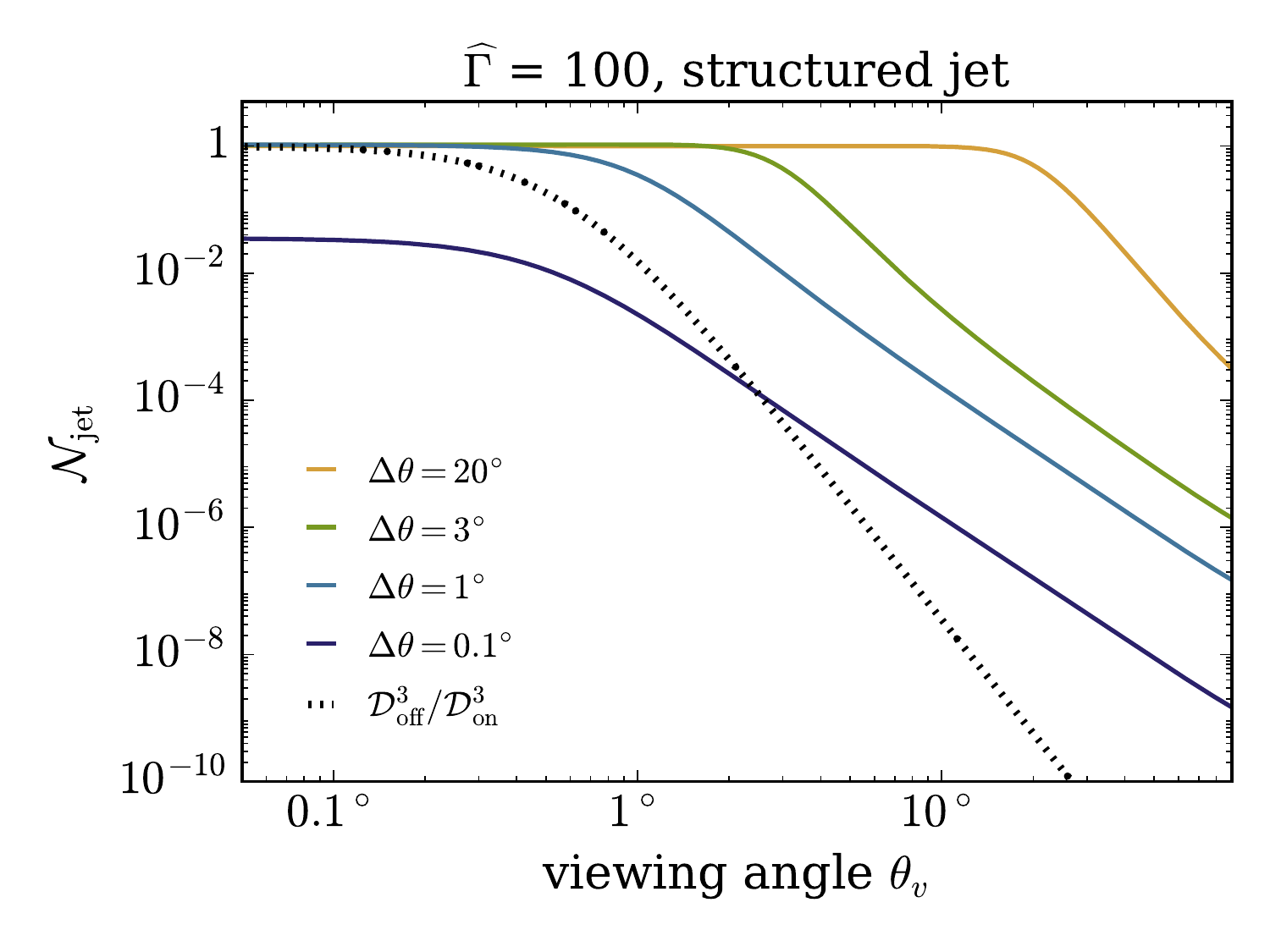}\hfill\includegraphics[width=0.5\linewidth]{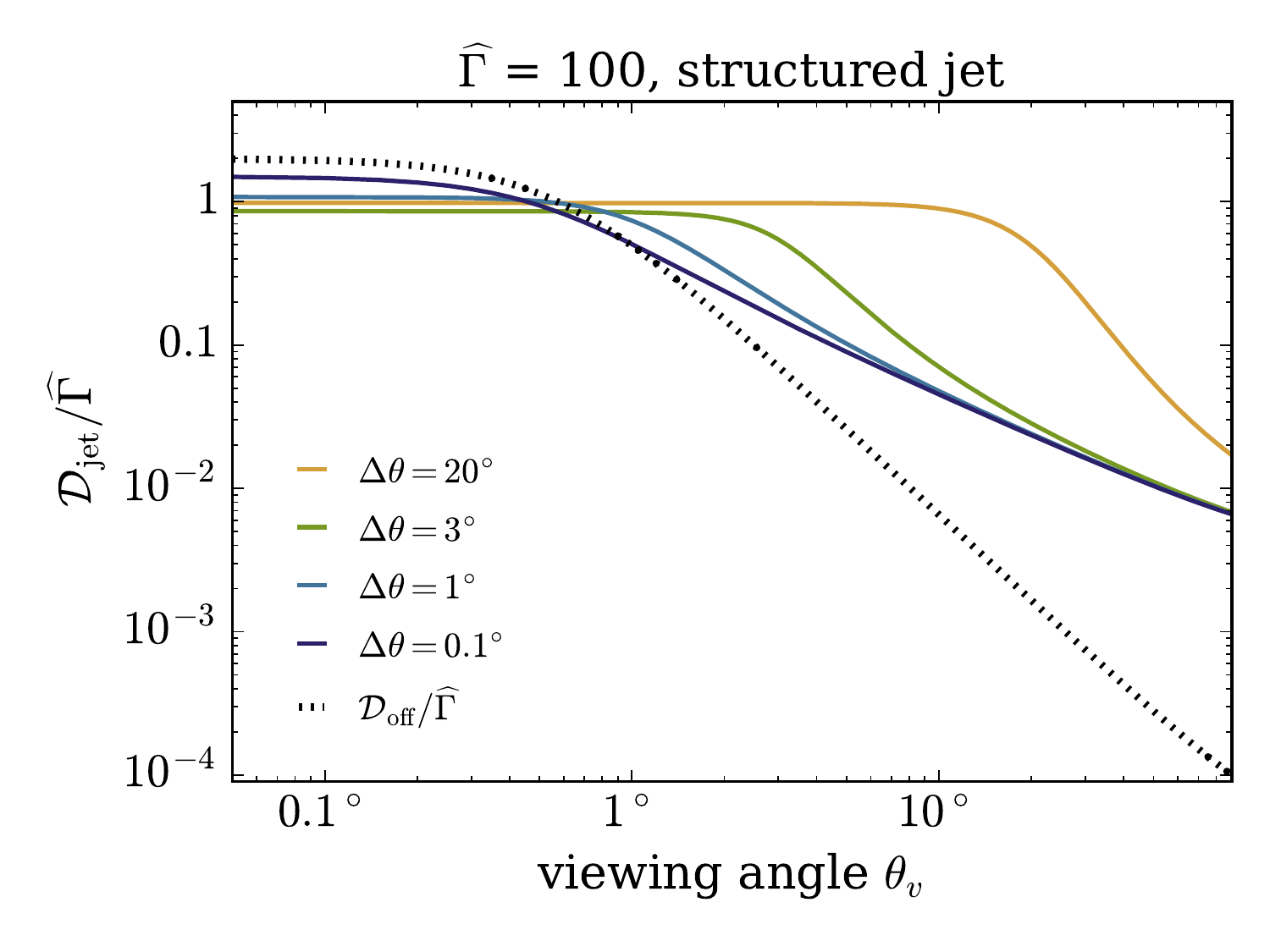}\\[-0.2cm]
\caption[]{The scaling factor $\mathcal{N}_{\rm jet}$ ({\it left}) defined in Eq.~(\ref{eq:Njet}) and effective Doppler factor $\mathcal{D}_{\rm jet}$ ({\it right}) defined in Eq.~(\ref{eq:Djet}) for a top-hat jet ({\it top}) and a structured jet ({\it bottom}). The dotted lines in the plots indicate the expected scaling of $\mathcal{N}_{\rm jet}$ and $\mathcal{D}_{\rm jet}/\widehat{\Gamma}$ for a top-hat jet observed at a large viewing angle $\theta_v$.}\label{fig2}
\end{figure*}
%%%%%%%%%%%%%%%%%%%%%%%%%%%%

\section{Jet Structure and Off-Axis Scaling}\label{sec2}

The previous discussion simplifies if we can consider an emission region that moves at a constant velocity $\boldsymbol\beta$. The energy fluence $\mathcal{F}$ in the observer's rest frame is then related to the bolometric energy $E'$ in the source rest frame as~\citep{Granot:2002za}
\begin{equation}\label{eq:plasmoid}
\mathcal{F} = \frac{1+z}{4\pi d_L^2} \mathcal{D}^3 E'\,,
\end{equation}
with $\mathcal{D} = \left[\Gamma(1-{\boldsymbol\beta}\cdot{\bf n}_{\rm obs})\right]^{-1}$. This approximates the case of a thin GRB jet observed at large viewing angle, $\theta_v\gg\Delta\theta$ and $\theta_v\gg1/\Gamma$. In this case expression (\ref{eq:plasmoid}) allows to estimate the off-axis emission from on-axis predictions by re-scaling the particle fluence $F$ (units of ${\rm GeV}^{-1}$ ${\rm cm}^{-2}$) by a factor $\eta = \mathcal{D}_{\rm off}/\mathcal{D}_{\rm on}$ as 
\begin{equation}\label{eq:naivescaling}
F_{\rm off}(\epsilon) = \eta F_{\rm on}(\epsilon/\eta)\,,
\end{equation}
where $\mathcal{D}_{\rm off}$ accounts for the viewing angle with respect to the jet boundary. This simple approximation was chosen by~\cite{ANTARES:2017bia} to account for the off-axis scaling of on-axis neutrino fluence predictions by~\cite{Kimura:2017kan} for the case of GRB 170817A. However, this scaling approach can only be considered to be a first order approximation and does not capture all relativistic effects, including intermediate situations where the kinematic angle $1/\Gamma$ becomes comparable to the viewing angle or jet opening angle or more complex situations of structured jets (see also the discussion by \cite{Biehl:2017qen}). 

In the following we will derive a generalization of the naive scaling relation (\ref{eq:naivescaling}) that applies to a larger class of jet structures and relative viewing angles. Expression (\ref{eq:fluencefinal}) derived in the previous section relates the observed fluence in photons or neutrinos to their internal energy density in the rest frame of the shell. The distribution of total energy and Lorentz factor with respect to the solid angle $\Omega^*$ is determined by the physics of the central engine and its interaction with the remnant progenitor environment before the jet emerges. It is therefore convenient to rewrite Eq.~(\ref{eq:fluencefinal}) in terms of a bolometric energy per solid angle in the GRB's rest frame~\citep{Salafia:2015vla},
\begin{equation}\label{eq:fluence} 
\mathcal{F} = \frac{1+z}{4\pi d_L^2}\int {\rm d}\Omega^* \frac{\mathcal{D}^3(\Omega^*)}{\Gamma(\theta^*)} \frac{{\rm d}E^*}{{\rm d}\Omega^*}\,.
\end{equation}
Using the relation ${{\rm d}E^*}/{{\rm d}\Omega^*} = \Gamma{{\rm d}E'}/{{\rm d}\Omega^*}$, one can recognize Eq.~(\ref{eq:fluence}) as the natural extension of Eq.~(\ref{eq:plasmoid}) for a spherical distribution of emitters. 
We can identify the angular distribution of internal energy from Eq.~(\ref{eq:fluencefinal}) as
\begin{equation}
\frac{{\rm d}E^*}{{\rm d}\Omega^*} = \frac{1}{4\pi} cT_{\rm GRB} 4\pi r_{\rm dis}^2(\theta^*)\Gamma^2(\theta^*)u'(\theta^*)\,.
\end{equation}
The jet structures that we are going to investigate in the following are parametrized in terms of the angular dependence of the Lorentz factor $\Gamma(\theta^*)$ and the kinetic energy ${\rm d}E^*/{\rm d}\Omega^*$ in the engine's rest frame. We will consider two cases:\\
\noindent {\it (i)} a {\it top-hat} (uniform) jet with
\begin{equation}\label{eq:E1}
\frac{{\rm d}E^*}{{\rm d}\Omega^*} = \frac{\widehat E}{4\pi}\Theta(\Delta\theta-\theta^*)\,,
\end{equation}
and
\begin{equation}\label{eq:G1}
\Gamma(\theta^*) = 1 + (\widehat{\Gamma}-1)\Theta(\Delta\theta-\theta^*)\,,
\end{equation}
corresponding to a constant Lorentz factor $\widehat\Gamma$ within a half-opening angle $\Delta\theta$ and\\[0.3cm]
\noindent {\it (ii)} a {\it structured} jet with
\begin{equation}\label{eq:E2}
\frac{{\rm d}E^*}{{\rm d}\Omega^*} = \frac{\widehat E}{4\pi}\frac{1}{1+(\theta^*/\Delta\theta)^{s_1}}\,,
\end{equation}
and
\begin{equation}\label{eq:G2}
\Gamma(\theta^*) = 1 + \frac{\widehat\Gamma-1}{1+(\theta^*/\Delta\theta)^{s_2}}\,.
\end{equation}
Both jet models are normalized to the core energy $\widehat E$ and Lorentz factor $\widehat \Gamma$ at the jet core. We use $s_1=5.5$ and $s_2=3.5$ in the following, corresponding to the best-fit parameters for the afterglow emission of GRB~170817A~\citep{Ghirlanda:2018uyx}. Note that in the limit $s_1\to \infty$ and $s_2\to\infty$, the structured jet is identical to the top-hat jet.

With these two jet models, we can now study the generalized off-axis scaling of the fluence (\ref{eq:fluence}). It is convenient to express the energy fluence~(\ref{eq:fluence}) in a form similar to the special case~(\ref{eq:plasmoid}) as
\begin{equation}\label{eq:fluencenew} 
\mathcal{F} = \frac{1+z}{4\pi d_L^2}\mathcal{N}_{\rm jet}\widehat{E}\,,
\end{equation}
where we introduce the jet scaling factor
\begin{equation}\label{eq:Njet}
\mathcal{N}_{\rm jet}(\theta_{v}) \equiv \int {\rm d}\Omega^* \frac{\mathcal{D}^3(\Omega^*)}{\Gamma(\theta^*)} \frac{1}{\widehat{E}}\frac{{\rm d}E^*}{{\rm d}\Omega^*}\,.
\end{equation}
The top left panel of Fig.~\ref{fig2} show this normalization factor for a top-hat jet for a variable viewing angle. The asymptotic behavior of the top-hat jet can be easily understood: For an on-axis observer with $\theta_v\ll\Delta\theta$ and jet factor approaches a constant. For high Lorentz factors, $\Gamma\Delta\theta\gg 1$, the emission from the edge of the jet is subdominant and the jet factor reaches $\mathcal{N}_{\rm jet}\simeq1$. The core energy $\widehat E$ is in this case equivalent to the isotropic-equivalent energy in the observer's frame. For low Lorentz factors, $\Gamma\Delta\theta\ll 1$, the edge of the jet becomes visible and the jet factor becomes $\mathcal{N}_{\rm jet}\simeq 2(\Gamma\Delta\theta)^2$.  On the other hand, for off-axis emission with $\theta_v\gg\Delta\theta$ the jet factor reproduces the expected $\mathcal{D}_{\rm off}^3$-scaling of Eq.~(\ref{eq:plasmoid}). The bolometric energy in the GRB and jet frame are related as $E^* = \Gamma E' \simeq (\Delta\Omega_{\rm jet}/4\pi)\widehat{E}$. For comparison, we show in the upper plot in Fig.~\ref{fig1} the naive scaling $(\mathcal{D}_{\rm off}/\mathcal{D}_{\rm on})^3$ expected from Eq.~(\ref{eq:plasmoid}), not correcting for the jet opening angle.

The case of a structured jet is shown in the bottom left panel of Fig.~\ref{fig2}. Similar to the case of the top-hat jet, at small viewing angles, $\theta_v\ll\Delta\theta$, the jet factor is independent of the viewing angle and reaches $\mathcal{N}_{\rm jet}\simeq1$ if $\Gamma\Delta\theta\gg 1$. However, the behavior at a large viewing angle, $\theta_v\gg\Delta\theta$, becomes more complex. The scaling with $\theta_v$ is much shallower than $(\mathcal{D}_{\rm off}/\mathcal{D}_{\rm on})^3$ expected from Eq.~(\ref{eq:plasmoid}) and a top-hat jet. 

We can extend the scaling of the energy fluence (\ref{eq:fluencenew}) to that of the particle fluence $F$. The particle fluence observed at an energy $\epsilon$ is related to contributions across the shell at energy $\epsilon' = \epsilon(1+z)/\mathcal{D}$. The particle fluence $F$ (in units of ${\rm GeV}^{-1} {\rm cm}^{-2}$) can then be derived following the same line of arguments used for the energy fluence and can be expressed as:
\begin{equation}\label{eq:generalspectrum}
\epsilon^2F(\epsilon) = \frac{1+z}{4\pi d_L^2}\int {\rm d}\Omega^* \frac{\mathcal{D}^3(\Omega^*)}{\Gamma(\theta^*)}\frac{{\rm d}E^*}{{\rm d}\Omega^*}\left[\frac{\epsilon'^2n'(\theta^*,\epsilon')}{u'(\theta^*)}\right]_{\epsilon' = \epsilon\frac{1+z}{\mathcal{D}(\Omega^*)}}\,.
\end{equation}
In the following we will assume that the internal spectrum only mildly varies across the sub-shell, $n'(\theta^*,\epsilon')/u'(\theta^*) \simeq n'(\epsilon')/u'$. We can then find an approximate solution to Eq.~(\ref{eq:generalspectrum}) of the form
\begin{equation}\label{eq:approximation}
\epsilon^2{F}(\epsilon) \simeq \frac{1+z}{4\pi d_L^2}\mathcal{N}_{\rm jet}\widehat{E} \left[\frac{\epsilon'^2n'(\epsilon')}{u'}\right]_{\epsilon' = \epsilon\frac{1+z}{\mathcal{D}_{\rm jet}}}\,,
\end{equation}
where we define the average Doppler boost as
\begin{equation}\label{eq:Djet}
\mathcal{D}_{\rm jet}(\theta_{v})\equiv \int {\rm d}\Omega^*\frac{\mathcal{D}^3(\Omega^*)}{\Gamma(\theta^*)}\frac{{\rm d}E^*}{{\rm d}\Omega^*}\bigg/\int {\rm d}\Omega^*\frac{\mathcal{D}^2(\Omega^*)}{\Gamma(\theta^*)}\frac{{\rm d}E^*}{{\rm d}\Omega^*}\,.
\end{equation}
Note that, by design, approximation (\ref{eq:approximation}) conserves the total energy {\it and} particle fluence from the source.

The right panels of Figure~\ref{fig2} show the normalized average Doppler factor $\mathcal{D}_{\rm jet}/\widehat{\Gamma}$ for the top-hat jet ({\it top}) and the structured jet ({\it bottom}). For on-axis observation, $\theta_v\ll\Delta \theta$, the average Doppler factor becomes independent of viewing angle. For high Lorentz factors, $\Gamma\Delta\theta\gg 1$, it approaches the Lorentz factor in the jet center, $\mathcal{D}_{\rm jet} \simeq \widehat{\Gamma}$. Only for narrow jets, $\Gamma\Delta\theta\ll 1$, and on-axis views we approach the on-axis Doppler limit $\mathcal{D}_{\rm jet} \simeq 2\widehat{\Gamma}$.

Again, the off-axis emission, $\theta_v\gg\Delta \theta$, shows quite different asymptotic behaviors for the two jet structures. In the case of the top-hat jet ({\it top}) the average Doppler factor approaches the naive scaling with off-axis Doppler factor $\mathcal{D}_{\rm off}$. Narrow top-hat jets, $\Delta\theta\Gamma\ll 1$, can be well approximated by $\mathcal{D}_{\rm jet}\simeq\mathcal{D}_{\rm off}$ over the full range of viewing angles. For the structured jet ({\it bottom}) the scaling of $\mathcal{D}_{\rm jet}$ with large viewing angle does not follow the naive $\mathcal{D}_{\rm off}$ scaling and lead to significantly higher Doppler factor.

With the quantities $\mathcal{N}_{\rm jet}$ and $\mathcal{D}_{\rm jet}$ we can now provide a revised scaling relation for the off-axis particle fluence $F$ (units of ${\rm GeV}^{-1} {\rm cm}^{-2}$): 
\begin{equation}\label{eq:betterscaling}
F_{\rm off}(\epsilon) \simeq \frac{\mathcal{N}_{\rm jet}(\theta_v)}{\mathcal{N}_{\rm jet}(0)}\frac{1}{\eta^2}F_{\rm on}(\epsilon/\eta)\,.
\end{equation} 
Here, we define in analogy to Eq.~(\ref{eq:naivescaling}) $\eta = {\mathcal{D}_{\rm jet} (\theta_v)}/{\mathcal{D}_{\rm jet}(0)}$, but in terms of the average Doppler factor in Eq.~(\ref{eq:Djet}) for different observer locations. Many GRB calculations are based on the assumption of an on-axis observer of a uniform jet with wide opening angle. In this case the on-axis calculation is based on $\mathcal{N}_{\rm jet}(0)\simeq1$ and $\mathcal{D}_{\rm jet}(0)\simeq\Gamma$, as can be seen in the top plots of Fig.~\ref{fig2}. Note that for top-hat jets observed at a large viewing angle, $\theta_v\gg\Delta \theta$, the ratio of the jet factors approaches ${\mathcal{N}_{\rm jet}(\theta_v)}/{\mathcal{N}_{\rm jet}(0)} \simeq \eta^3$ ({\it cf.}~top left panel of Fig.~\ref{fig2}) and in this case Eq.~(\ref{eq:betterscaling}) reproduces the naive scaling relation (\ref{eq:naivescaling}).

In principle, the scaling relation (\ref{eq:betterscaling}) applies to photon and neutrino predictions based on arbitrary jet structures and viewing angles. However, a crucial underlying assumption of the approximation (\ref{eq:betterscaling}) is that the relative emission spectrum only mildly varies across the sub-shell, $n'(\theta^*,\epsilon')/u'(\theta^*) \simeq n'(\epsilon')/u'$. We will see in the following that the emissivity of structured jets in the internal shock model can have strong local variations of magnetic fields and photon densities across the shock, which can jeopardize this condition. In this case, the calculation needs to be carried out using the exact expression (\ref{eq:generalspectrum}).

%%%%%%%%%%%%%%%%%%%%%%%%
\begin{figure}\centering
\includegraphics[width=\linewidth]{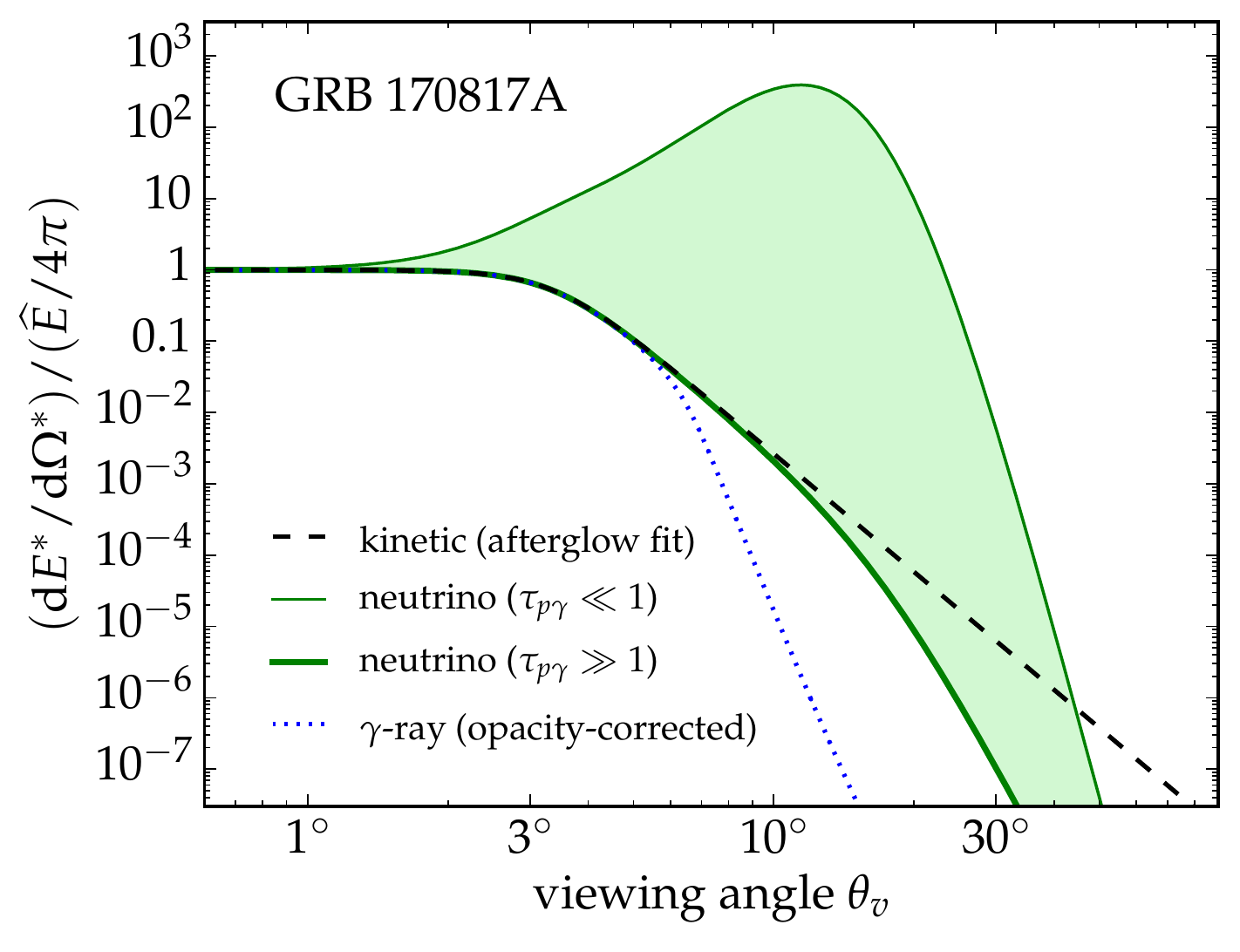}\\[-0.2cm]
\caption[]{Relative angular distribution of the energy associated with the bulk flow (solid black line), neutrinos at low and high opacity (thin \& thick green line), and $\gamma$-rays corrected for Thomson scattering in the shell (dotted blue line).}\label{fig3}
\end{figure}
%%%%%%%%%%%%%%%%%%%%%%%%

\section{Neutrino Fluence from Structured Jets}\label{sec3}

Internal shocks from colliding sub-shells of the GRB engine are expected to accelerate protons (and heavier nuclei) entrained in the GRB outflow. The spectrum of cosmic ray protons is assumed to follow a power-law close to $\epsilon_p^{-2}$ up to an effective cutoff that is determined by the relative efficiency of cosmic ray acceleration and competing energy loss processes. Based on the $\gamma$-ray fluence of the burst, one can estimate the internal energy densities of cosmic rays, photons and magnetic fields. The internal photon density allows to predict the opacity of individual sub-shells to proton-photon ($p\gamma$) interactions. Neutrino production follows predominantly from the production of pions, that decay via $\pi^+\to\mu^+\nu_\mu$ followed by $\mu^+\to e^+\nu_e\bar\nu_\mu$ or the charge-conjugate processes. The presence of strong internal magnetic fields leads to synchrotron loss of the initial protons and secondary charged particles before their decay. The mechanism was initially introduced by \cite{Waxman:1997ti} for the case of an on-axis jet with wide opening angle and has been studied in variations by several authors since~\citep{Guetta:2003wi,Murase:2005hy,Anchordoqui:2007tn,Ahlers:2011jj,He:2012tq,Zhang:2012qy,Tamborra:2015qza,Denton:2017jwk}.

The energy densities of photons, magnetic fields, and cosmic rays are limited by the efficiency of internal collisions (IC) of merging sub-shells to convert bulk kinetic energy of the flow into total internal energy of the merged shell. In the rest frame of the central engine, we parametrize the total internal energy from the kinetic energy of the outflow via an angular-dependent efficiency factor $\eta_{\rm IC}$ as
\begin{equation}\label{eq:EIS}
\frac{{\rm d}E^*_{\rm IC}}{{\rm d}\Omega^*}=
\eta_{\rm IC}(\theta^*)\frac{{\rm d}E^*}{{\rm d}\Omega^*}\,.
\end{equation}
To first order, the efficiency of converting bulk kinetic energy into internal energy can be estimated by energy and momentum conservation~\citep{Kobayashi:1997jk}. In Appendix~\ref{appA} we introduce a simple model of the efficiency factor as a function of the Lorentz factor $\Gamma(\theta^*)$ and the asymptotic efficiency $\eta_\infty$ for large Lorentz factors. The partition of the internal energy into $\gamma$-rays, cosmic rays and magnetic fields is then parametrized as
\begin{equation}\label{eq:Einternal}
\frac{{\rm d}E_x^*}{{\rm d}\Omega^*}=
\varepsilon_x\frac{{\rm d}E^*_{\rm IC}}{{\rm d}\Omega^*}\,,
\end{equation}
with the corresponding energy fraction $\varepsilon_\gamma$, $\varepsilon_p$ and $\varepsilon_B$, respectively.

Using relation (\ref{eq:fluencefinal}), we can express the internal photon energy density as
\begin{equation}
u_\gamma'(\theta^*) \simeq \frac{L_\gamma^{\rm iso}/\mathcal{N}_{\rm jet}(\theta_v)}{c r_{\rm dis}^2(\theta^*)\Gamma^2(\theta^*)}\frac{1}{\widehat E}_\gamma\frac{{\rm d}E_\gamma^*}{{\rm d}\Omega^*}\,,
\end{equation}
where the isotropic-equivalent luminosity is defined by the $\gamma-$ray fluence as $L^{\rm iso}_\gamma \equiv 4\pi d_L^2\mathcal{F}_\gamma/T_{90}$. Neutrino production from $p\gamma$ interactions is determined by the opacity $\tau_{p\gamma} \simeq ct'_{\rm dyn}\sigma_{p\gamma}n'_\gamma$ of individual merging sub-shells. If we relate the shell position and width to the variability of the central engine (see Fig.~\ref{fig1}) and assume that the $\gamma$-ray spectrum is observed at a peak photon energy $\epsilon_{\rm peak}$, we can express the $p\gamma$ opacity as
\begin{equation}\label{eq:taupgamma}
\tau_{p\gamma}(\theta^*) \simeq \frac{\sigma_{p\gamma}L^{\rm iso}_\gamma}{c^2\Delta t_{\rm eng}\epsilon_{\rm peak}}\frac{\mathcal{D}_{\rm jet}(\theta_v)}{\mathcal{N}_{\rm jet}(\theta_v)}\frac{1}{\Gamma^5(\theta^*)}\frac{1}{\widehat{E}_\gamma}\frac{{\rm d}E_\gamma^*}{{\rm d}\Omega^*}\,.
\end{equation}
For an on-axis observer of a wide ($\Gamma\Delta\theta\gg1$) jet this reduces to the familiar $\widehat{\Gamma}^{-4}$-scaling~\citep{Waxman:1997ti} since $\mathcal{D}_{\rm jet}(0)\simeq \Gamma$ and $\mathcal{N}_{\rm jet}(0)\simeq 1$ (see Fig.~\ref{fig2}). Note that the opacity is independent of viewing angle; the appearance of the quantities $\mathcal{N}_{\rm jet}$ and $\mathcal{D}_{\rm jet}$, that strongly depend on jet structure and viewing angle, compensate the corresponding scaling of the peak emission energy $\epsilon_{\rm peak}$ and isotropic-equivalent luminosity $L^{\rm iso}_\gamma$ in the observer's frame. We can finally approximate the neutrino scaling with the jet angle as
\begin{equation}\label{eq:Enu}
\frac{{\rm d}E_\nu^*}{{\rm d}\Omega^*}
\simeq\frac{3}{4}K_\pi\frac{\varepsilon_p}{\varepsilon_\gamma}\left(1-e^{-\kappa\tau_{p\gamma}(\theta^*)}\right)\frac{{\rm d}E_\gamma^*}{{\rm d}\Omega^*}\,.
\end{equation}
Here we account for the inelasticity $\kappa\simeq 0.2$ of photo-hadronic interactions. The pre-factors in Eq.~(\ref{eq:Enu}) accounts for the fraction of charged-to-neutral pions, $K_\pi\simeq 1/2$, and for three neutrinos carrying about 1/4th of the pion energy. The combination $\varepsilon_p/\varepsilon_\gamma$ corresponds to the non-thermal baryonic loading factor $\xi_p$, which we fix at $\xi_p\simeq1$ in the following. 

The formalism outlined here so far follows the standard approach of neutrino production in the internal shock model. The new aspect that we want to highlight is the angular distribution of total neutrino energy (\ref{eq:Enu}) in structure jet models. Depending on the opacity of the shell the neutrino scaling with jet angle becomes
\begin{equation}
\frac{{\rm d}E_\nu^*}{{\rm d}\Omega^*}\propto\begin{cases}\frac{{\rm d}E_{\rm IC}^*}{{\rm d}\Omega^*}&\tau_{p\gamma} \gg 1\,,\\
\frac{1}{\Gamma^5(\theta^*)}\left(\frac{{\rm d}E_{\rm IC}^*}{{\rm d}\Omega^*}\right)^2&\tau_{p\gamma} \ll 1\,.\end{cases}
\end{equation}
The strong angular dependence of neutrino emission in low opacity regions can have a significant influence on the neutrino predictions, as we will illustrate by the case of GRB 170817A in the following.

\section{Prompt Emission of GRB 170817A}\label{sec4}

As an illustration of neutrino production in structured jets we will discuss the prompt emission of the recent short GRB 170817A observed in coincidence with the gravitational wave GW170817 from a binary neutron star merger~\citep{GBM:2017lvd,Monitor:2017mdv}. The spectrum observed with Fermi-GBM is best described as a Comptonized spectrum, $n_\gamma(\epsilon) \propto \epsilon^\alpha\exp(-{(2+\alpha)\epsilon}/{\epsilon_{\rm peak}})$, with spectral index $\alpha\simeq0.14\pm0.59$ and peak photon energy $\epsilon_{\rm peak}\simeq(215\pm54)$~keV~\citep{Goldstein:2017mmi}. The energy fluence integrated in the 10--1000~keV range is $\mathcal{F}_\gamma\simeq (1.4\pm0.3)\times10^{-7}{\rm erg}\,{\rm cm}^{-2}$. The source is located at a luminosity distance of $d_L\simeq 41$~Mpc corresponding to a redshift $z\simeq 0.01$. The variability of the central engine is $t_{\rm var}\simeq0.125$s with an emission time of $T_{90}\simeq 2$s. From this we can calculate the isotropic-equivalent energy as $E^{\rm iso}_\gamma \simeq (2.8\pm0.6)\times 10^{46}$~erg.
 
Based on afterglow emission, \cite{Ghirlanda:2018uyx} derived a model for the angular dependence of the kinetic energy of the outflow, based on the parametrizations of Eqs.~(\ref{eq:E2}) and (\ref{eq:G2}) with best-fit parameters $s_1=5.5$ and $s_2=3.5$, opening angle $\Delta\theta\simeq 3.4^\circ$, core Lorentz factor $\widehat{\Gamma}\simeq250$, core energy $\widehat{E}\simeq 2.5\times10^{52}$~erg and viewing angle $\theta_v\simeq15^\circ$. Alternative models of the outflows have been presented by \cite{Lazzati:2017zsj}, \cite{Troja:2018ruz}, \cite{Margutti:2018xqd}, \cite{Lamb:2018qfn} or \cite{Lyman:2018qjg}.

\subsection{Gamma-Ray Fluence}

Before we turn to the neutrino fluence, it is illustrative to compare the structured jet model of~\cite{Ghirlanda:2018uyx} based on afterglow observations to the expected prompt $\gamma$-ray emission from the internal shock model. For this comparison it is crucial to account for angular-dependent internal photon absorption. The opacity of individual sub-shells with respect to Thompson scattering on baryonic electrons is given as $\tau_T \simeq ct'_{\rm dyn}\sigma_Tn'_e$ with Thomson cross section $\sigma_T\simeq 0.67$~barn and local baryonic electron density
\begin{equation}
n'_e \simeq \frac{Y_e}{cr^2_{\rm dis}m_p\Gamma(\theta^*)}\left[
\frac{1}{T_{\rm GRB}(\Gamma(\theta^*)-1)}\frac{{\rm d}E^*}{{\rm d}\Omega^*}\right]\,.
\end{equation}
The term in square brackets correspond to the angular-dependent mass flow ${\rm d}\dot{M}/{\rm d}\Omega^*$ of the structured jet. For the proton fraction of the flow we assume $Y_e\simeq 1/2$ in the following. We can then account for $\gamma$-ray absorption by Thomson scattering as
\begin{equation}\label{eq:EGRB}
\frac{{\rm d}E_{\rm GRB}^*}{{\rm d}\Omega^*} \simeq  \frac{1-e^{-\tau_T(\theta^*)}}{\tau_T(\theta^*)}\frac{{\rm d}E_\gamma^*}{{\rm d}\Omega^*}\,.
\end{equation}
Figure~\ref{fig3} shows this angular distribution of emitted $\gamma$-rays as a dotted blue line.

From this $\gamma$-ray emission model, we calculate a jet scaling factor $\mathcal{N}_{\rm jet}\simeq 1.4\times10^{-5}$ for a viewing angle $\theta_v\simeq15^\circ$. Following Eq.~(\ref{eq:fluencenew}), the internal $\gamma$-ray energy at the jet core is therefore required to reach $\widehat{E}_\gamma \simeq (2.1\pm0.4)\times10^{51}$~erg to be consistent with the fluence level observed by Fermi-GBM. Assuming an asymptotic efficiency factor $\eta_{\infty} \simeq  0.2$ in Eq.~(\ref{eq:EIS}) we can estimate the total internal energy of the sub-shell at the jet center as $\widehat{E}_{\rm IC}\simeq 5\times10^{51}~{\rm erg}$. This is consistent with the $\gamma$-ray observation if we require that an energy fraction of $\varepsilon_\gamma\simeq 0.41\pm0.09$ contributes to the $\gamma$-ray emission of the burst. 

For the prediction of the corresponding neutrino fluence we have to make an assumption about the relative photon target spectrum $n_\gamma'(\theta^*,\epsilon')/u'_\gamma(\theta^*)$ at angular distance $\theta^*$ in the sub-shell. In general, we don't expect that the spectral features remain constant across the shell, owing to the strong local variations from synchrotron loss in magnetic fields and photon absorption via Thomson scattering. Indeed, the $\gamma$-ray emissivity at an assumed viewing angle of $\theta_v\simeq15^\circ$ is strongly suppressed by the opacity of the photosphere; {\it cf.}~Fig.~\ref{fig3}. To study the dependence of our neutrino fluence predictions on this model uncertainty we consider two scenarios. In both cases we assume that the internal photon spectrum follows a Comptonized spectrum with low-energy index $\alpha=0.14$. For the exponential cutoff we assume:\\
\noindent {\it (a)} a {\it constant} co-moving peak 
\begin{equation}\label{eq:fixed1}
\epsilon_{\rm peak}'(\theta^*) \simeq \frac{215{\rm keV}}{\mathcal{D}_{\rm jet}} \simeq 75{\rm keV}\,,
\end{equation}
where $\mathcal{D}_{\rm jet}\simeq2.9$ is the average Doppler factor (\ref{eq:Djet}) based on the angular-dependent $\gamma$-ray emission (\ref{eq:EGRB}) and\\
\noindent {\it (b)} a {\it scaled} co-moving peak 
\begin{equation}\label{eq:fixed2}
\epsilon_{\rm peak}'(\theta^*) \simeq 75{\rm keV}\frac{\Gamma(\theta_v)}{\Gamma(\theta^*)}\,,
\end{equation}
which corresponds to a fixed peak position $\epsilon^*_{\rm peak} = \Gamma\epsilon'_{\rm peak}$ in the rest frame of the central engine. 

These two models are chosen such that the internal $\gamma$-ray emissivity is consistent with the $\gamma$-ray spectrum of GRB 170817A observed by Fermi-GBM at a viewing angle of $15^\circ$. However, note that model {\it (a)} implies that the peak photon energy for the {\it on-axis} observations would reach energies $\epsilon_{\rm peak}\simeq20$~MeV, in tension with the peak distribution inferred from GRBs observed by Fermi-GBM~\citep{Gruber:2014iza}. The phenomenological model {\it (b)} is motivated by the discussion of \cite{Ioka:2019jlj}, who study the consistency of the on-axis emission of GRB 170817A with the $E_\gamma^{\rm iso}$-$\epsilon_{\rm peak}$ correlation suggested by~\cite{Amati:2006ky}. Here, the on-axis fluence is expected to peak at $\epsilon_{\rm peak}\simeq178$~keV.

%%%%%%%%%%%%%%%%%%%%%%%%
\begin{figure}\centering
\includegraphics[width=\linewidth]{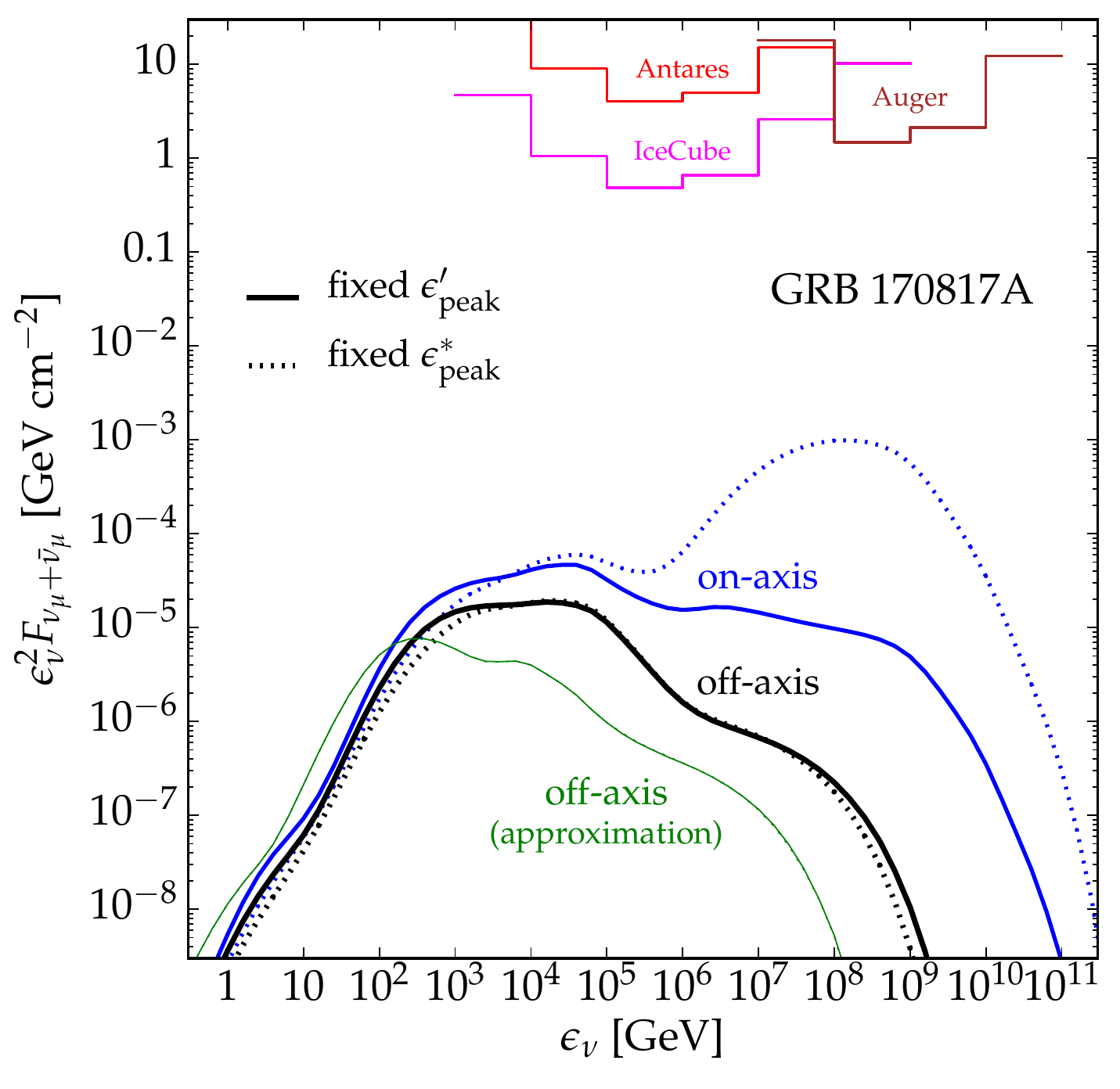}\\[-0.2cm]
\caption[]{Predicted fluence of muon neutrinos ($\nu_\mu+\bar\nu_\mu$) associated with the prompt emission in the best-fit structured jet model of \cite{Ghirlanda:2018uyx}. We show the predictions based on a fixed photon peak in the shell frame (``fixed $\epsilon'_{\rm peak}$'', solid lines) using Eq.~(\ref{eq:fixed1}) and in the engine frame (``fixed $\epsilon^*_{\rm peak}$'', dotted lines) using Eq.~(\ref{eq:fixed2}). The thick black lines show the off-axis emission at a viewing angle $\theta_v=15^\circ$. The blue lines show the corresponding prediction for the on-axis emission, which has a strong dependence on the internal photon spectrum. The thin green lines show the result of an approximation based on the standard on-axis calculation of uniform jets \citep{Waxman:1997ti} with jet parameters from the structured jet model at $\theta^*=\theta_v$. The upper solid lines indicate the 90\% C.L.~upper limit on the fluence from \cite{ANTARES:2017bia}.}\label{fig4}
\end{figure}
%%%%%%%%%%%%%%%%%%%%%%%%

\subsection{Neutrino Fluence}

As we discussed in section~\ref{sec3}, the neutrino emissivity of a structured jet is expected to deviate from the angular distribution of the observable $\gamma$-ray emission. For high opacity ($\tau_{p\gamma}\gg1$) regions of the shell the angular distribution of the neutrino emission is expected to follow the distribution of internal energy (\ref{eq:EIS}) that takes into account the efficiency of dissipation in internal collisions. This is shown for our efficiency model (\ref{eq:effIS}) as the thick green line in Fig.~\ref{fig4}. For low-opacity ($\tau_{p\gamma}\gg1$) regions, however, the energy distribution has an additional angular scaling from the opacity (\ref{eq:taupgamma}), as indicated by the thin green line. One can notice that a low opacity environment has an enhanced emission at jet angles $10^\circ$-$20^\circ$, which is comparable to our relative viewing angle. Note that the angular distributions in Fig.~\ref{fig3} are normalized to the value at the jet core and do not indicate the absolute emissivity of neutrinos or $\gamma$-rays, which depend on jet angle $\theta^*$ and co-moving cosmic ray energy $\epsilon'_{\rm p}$. 

At each jet angle $\theta^*$ we estimate the maximal cosmic ray energy based on a comparison of the acceleration rate to the combined rate of losses from synchrotron emission, $p\gamma$ interactions (Bethe-Heitler and photo-hadronic) and adiabatic losses. Our model predictions assume a magnetic energy ratio compared to $\gamma$-rays of $\xi_B = 0.1$ and a non-thermal baryonic loading of $\xi_p\simeq1$ (see Appendix~\ref{appB}). We calculate the neutrino emissivity $j'_{\nu_\alpha}(\theta^*,\epsilon'_\nu)$ from $p\gamma$ interactions with the photon background in sub-shells based on the Monte-Carlo generator \texttt{SOPHIA}~\citep{Mucke:1999yb}, that we modified to account for synchrotron losses of all secondary charged particles before their decay~\citep{Lipari:2007su}. The uncertainties regarding the photon target spectrum are estimated in the following via the two models {\it (a)} and {\it (b)} of the peak photon energy.
 
The expected fluence of muon neutrinos ($\nu_\mu + \bar\nu_\mu$) under different model assumptions is shown in Fig.~\ref{fig4}. The off-axis fluence at a viewing angle of $\theta_v\simeq15^\circ$ is indicated as thick black lines. The off-axis prediction has only a weak dependence on the angular scaling of the co-moving peak of the photon spectrum, Eqs.~(\ref{eq:fixed1}) or (\ref{eq:fixed2}), as indicated as solid and dotted lines, respectively. This is expected from the normalization of the model to the observed $\gamma$-ray fluence under this viewing angle. For comparison, we also show in Fig.~\ref{fig4} an approximation (thin green lines) of the off-axis neutrino fluence based on the on-axis top-hat jet calculation with Lorentz factor and neutrino emissivity evaluated at $\theta^*\simeq\theta_v$. This approximation has been used by~\cite{Biehl:2017qen} to scale the off-axis emission of the structured jet. Note that this approximation significantly underestimates the expected neutrino fluence of GRB 170717A compared to an exact calculation. 

Figure~\ref{fig4} also indicates the predicted neutrino fluence for an on-axis observer of the source located at the same luminosity distance. The extrapolated on-axis fluence shows a strong dependence on the model of the internal photon spectrum; model (\ref{eq:fixed2}) predicts a strong neutrino peak at the EeV scale that exceeds the prediction of model (\ref{eq:fixed1}) by two orders of magnitude. The relative difference of the neutrino fluence at the EeV scale follows from the ratio of $\epsilon'_{\rm peak}(0)$ for the two models~(\ref{eq:fixed1}) and (\ref{eq:fixed1}): For a fixed co-moving energy density of the shell, a lower peak photon energy corresponds to a higher photon density and also a higher threshold for neutrino production. One can also notice, that the on-axis neutrino fluence in the TeV range depends only marginally on the viewing angle. This energy scale is dominated by the emission of the jet at $\theta^*\simeq10^\circ-20^\circ$ and reflects the strong angular dependence of the neutrino emission in the rest frame of the central engine ({\it cf.}~Fig.~\ref{fig3}).

The upper thin solid lines in Fig.~\ref{fig4} show the 90\% confidence level (C.L.) upper limits on the neutrino flux of GRB 170817A from Antares, Auger and IceCube~\citep{ANTARES:2017bia}. The predicted neutrino fluence is orders of magnitude below these combined limits. However, our neutrino fluence predictions are proportional to the non-thermal baryonic loading factor, and we assume a moderate value of $\xi_p=1$ for our calculations. In any case, the predicted neutrino flux at an observation angle of $15^\circ$ is many orders of magnitude larger than the expectation from an off-axis observation of a uniform jet.

\section{Conclusions}\label{sec5}

In this paper, we have discussed the emission of neutrinos in the internal shock model of $\gamma$-ray bursts. The majority of previous predictions are based on the assumption of on-axis observations of uniform jets with wide opening angles. Here, we have extended the standard formalism of neutrino production in the internal shock model to account for arbitrary viewing angles and jet structures, parametrized by the angular distribution of kinetic energy and Lorentz factor of the outflow of the GRB engine.

One of the main results of this paper is a revised relation of the particle fluence between on- and off-axis observers given in Eq.~(\ref{eq:betterscaling}) based on the exact scaling of the energy fluence given by Eq.~(\ref{eq:Njet}) and an average Doppler factor defined by Eq.~(\ref{eq:betterscaling}). This relation allows to rescale previous on-axis calculations for off-axis observers, assuming that the relative emission spectrum is independent on the jet angle, as expected for uniform jets. The particle fluence under general conditions can be derived from the exact expression~(\ref{eq:generalspectrum}).

We have shown that the neutrino emissivity of structured jets can exhibit a strong relative dependence on the jet angle compared to the emission of $\gamma$-rays. We have illustrated this dependence for the case of GRB 170817A assuming a structured jet model inferred from afterglow observations. We have shown that this model is consistent with the observed $\gamma$-ray fluence if we take into account photon absorption at large jet angles. We find that the predicted off-axis neutrino emission at about $15^\circ$ is similar to the on-axis prediction in the TeV energy range and orders of magnitude larger than the expected fluence from an off-axis observation of a uniform jet. 

Neutrino fluence predictions in this paper followed from the standard internal shock model by \cite{Waxman:1997ti}, where the kinetic energy of the outflow is dissipated via colliding sub-shells. The size and location of the merging sub-shells is set by the variability of the central engine and the bulk Lorentz factor. Different mechanisms of dissipation, {\it e.g.}~magnetic reconnections as in the model by~\cite{Zhang:2012qy}, predict a different dissipation radius of the jet. Our results regarding the off-axis scaling of the emission derived in section~\ref{sec2} apply equally to this type of dissipation scenario.

Finally, variations of the dissipation radius in the continuous outflow of the GRB engine also allows for the possibility that cosmic rays interact with photons emitted from different locations along the jet axis. This additional photon background could become important in structured jets, where the relative motion of jet layers boosts the photon flux coming from an outer (inner) layer downstream (upstream) of the jet. This mechanism has been suggested to enhance the neutrino emission in blazar jets~\citep{Tavecchio:2014iza,Tavecchio:2014eia}. Two layers with Lorentz factors $\Gamma_{1/2}$ in the rest frame of the central engine have a relative Lorentz factor of $\Gamma_{\rm rel} = \Gamma_1\Gamma_2(1-\beta_1\beta_2)$. Assuming an extended structured jet with continuous emission along the jet axis, the cross-layer photon flux would be enhanced by a factor $\Gamma_{\rm rel}^2$. However, these assumptions are not suitable for transient sources and we will postpone a more detailed discussion of this effect to a future project.

\section*{Acknowledgements}

The authors acknowledge support by \textsc{Villum Fonden} under project no.~18994.

%%%%%%%%%%%%%%%%%%%%%%%%%%%%%%%%%%%%%%%%%%%%%%%%%%

\bibliographystyle{mnras}
\bibliography{references} % if your bibtex file is called example.bib
%%%%%%%%%%%%%%%%%%%%%%%%%%%%%%%%%%%%%%%%%%%%%%%%%%

%%%%%%%%%%%%%%%%% APPENDICES %%%%%%%%%%%%%%%%%%%%%

\appendix
\section{Efficiency of Internal Shocks}\label{appA}

Two sub-shells emitted from the central engine at times $t_1<t_2$ with Lorentz factors $\Gamma_1<\Gamma_2$ will eventually collide. The efficiency of converting bulk kinetic energy into internal energy $E'$ can be estimated by energy and momentum conservation~\citep{Kobayashi:1997jk}:
\begin{align}
E_{\rm tot}&=\Gamma_1M_1 + \Gamma_2M_2 = \Gamma (M+E')\,,\\
P_{\rm tot}&=\sqrt{\Gamma^2_1-1}M_1 + \sqrt{\Gamma^2_2-1}M_2 =\sqrt{\Gamma^2-1}(M+E')\,.
\end{align}
The efficiency of energy dissipation in internal collisions (IC) is then defined as 
\begin{equation}\label{eq:efficiency}
\eta_{\rm IC} = 1 - \frac{\Gamma M}{E_{\rm tot}}\,.
\end{equation}
In the relativistic limit, the Lorentz factor for the combined shells is
\begin{equation}
\Gamma \simeq \sqrt{\frac{\Gamma_1M_1 + \Gamma_2M_2}{M_1/\Gamma_1 + M_2/\Gamma_2}}\,.
\end{equation}
We will assume in the following that the variation of the central engine introduces variations in the energy of the form
\begin{equation}\label{eq:Gammavariation}
\sqrt{E_{\rm tot}^2-(\Gamma M)^2} \simeq x(\Gamma-1)M\,,
\end{equation}
with $x=\mathcal{O}(1)$. In the relativistic limit and equal-mass shells, $M_1\simeq M_2$, Eq.~(\ref{eq:Gammavariation}) becomes equivalent to a condition on the variation of the Lorentz factors, $|\Gamma_2-\Gamma_1|\simeq 2x\Gamma$, which is typically assumed in the internal shock model. On the other hand, Eq.~(\ref{eq:Gammavariation}) ensures that the efficiency approaches zero for slow outflows in the tail of structured jets. The two Eqs.~(\ref{eq:efficiency}) and (\ref{eq:Gammavariation}) define our model for the efficiency $\eta_{\rm IC}(\Gamma)$ of converting bulk kinetic energy to internal energy via colliding sub-shells. The efficiency rises with $\Gamma$ and approaches the asymptotic value $\eta_\infty = 1-1/\sqrt{1+x^2}$ at high Lorentz factors. We can express the efficiency in terms of the combined Lorentz factor and the asymptotic efficiency as
\begin{equation}\label{eq:effIS}
\eta_{\rm IC}(\Gamma) = 1 - \frac{\Gamma}{\sqrt{2\Gamma -1 + (\Gamma-1)^2/(1-\eta_\infty)^2}}\,.
\end{equation}
In this paper we will assume the asymptotic value $\eta_\infty\simeq0.2$ that corresponds to $x\simeq3/4$ in Eq.~(\ref{eq:Gammavariation}).

%%%%%%%%%%%%%%%%%%%%%%%%
\begin{figure}\centering
\includegraphics[width=\linewidth]{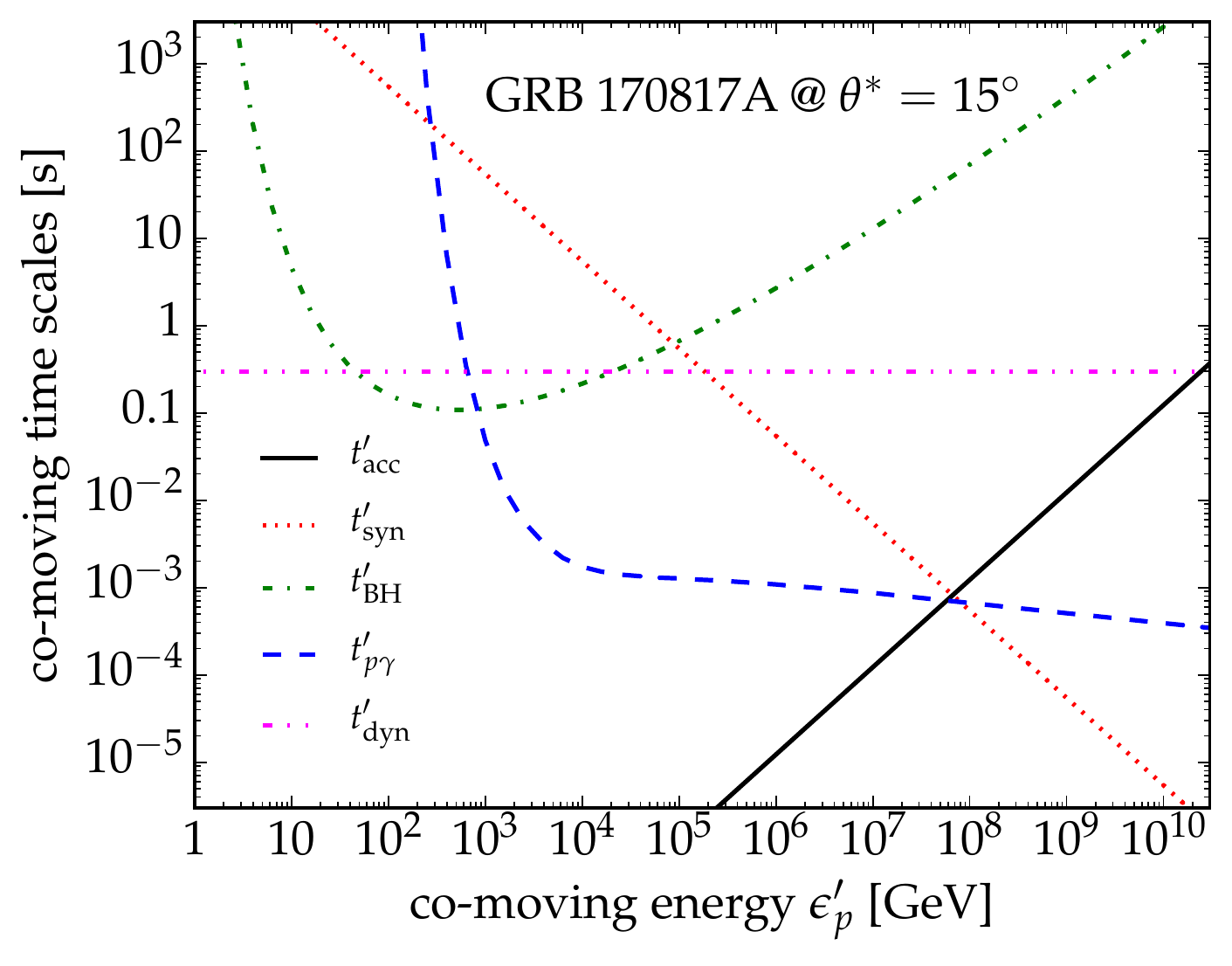}\\[-0.2cm]
\caption[]{The acceleration time scale (solid) in comparison to the time scale of synchrotron loss (dotted), Bethe-Heitler pair production (dotted-dashed), hadronic interactions (dashed) and adiabatic losses (double-dotted-dashed) at jet angle $\theta^*\simeq15^\circ$.}\label{fig5}
\end{figure}
%%%%%%%%%%%%%%%%%%%%%%%%

\section{Cosmic Ray Spectrum}\label{appB}

We assume that cosmic ray protons in the sub-shell follow an $\epsilon_p'^{-2}$ spectrum with an exponential cutoff $\epsilon'_{p,{\rm max}}$. The jet model determines the local magnetic field and spectral photon density at different jet angles $\theta^*$. Cosmic ray acceleration in internal shocks is expected to scale with the inverse of the Larmor radius or
\begin{equation}
t'^{-1}_{\rm acc} = \eta_{\rm acc}\frac{eB'}{\epsilon_p'}\,,
\end{equation}
where $\eta_{\rm acc}$ is the acceleration efficiency. In our calculation, we will assume high efficiencies of $\eta_{\rm acc}\simeq1$. The maximal cosmic ray energy can be determined by comparing the acceleration rate to the combined rate of losses:\\
\noindent {(i)} Adiabatic cooling of the expanding shell can be estimated by the the dynamical time scale of the central engine,
\begin{equation}
t'^{-1}_{\rm dyn} = \frac{1}{\Gamma\Delta t_{\rm eng}}\,.
\end{equation}
\noindent {(ii)} The angular-averaged synchrotron loss of cosmic ray protons in the magnetized shell is given as
\begin{equation}
t'^{-1}_{\rm syn} = \frac{e^4{B'}^2{\epsilon_p'}}{9\pi m_p^4}\,.
\end{equation}
\noindent {(iii)} The energy loss of $p\gamma$ interactions in the rest frame of the sub-shell is given by
\begin{align}\label{eq:gamma}
t'^{-1}_{p\gamma}
=\frac{\kappa}{2\gamma^2}\int{\rm d}\hat\epsilon\hat\epsilon\sigma_{p\gamma}(\hat\epsilon)\int\limits_{\hat\epsilon/2\gamma}\frac{{\rm d}x}{x^2}n'_\gamma(x)\,,
\end{align}
where $\kappa$ is the average inelasticity of the interaction with background photons and $\hat\epsilon=\epsilon_\gamma'\gamma(1-\cos\theta)$ the photon's energy in the rest frame of the proton with Lorentz boost $\gamma\simeq \epsilon_p'/m_p$.\\
\noindent {(iv)} Bethe-Heitler (BH) $e^+e^-$ pair production by cosmic ray scattering off background photons with time loss rate $t'^{-1}_{\rm BH}$ can be accounted for by the differential cross section calculated by \cite{Blumenthal:1970nn}. 

Our neutrino calculations are based on the Monte-Carlo generator \texttt{SOPHIA}~\citep{Mucke:1999yb}, that we modified to include synchrotron loss of {\it all} intermediate particles of the $p\gamma$-interaction cascade following~\cite{Lipari:2007su}. For a secondary particle with charge $Z$, mass $m$ and proper lifetime $\tau_0$, the ratio $x\equiv\epsilon'_f/\epsilon'_i$ of final to initial energy is following the probability distribution 
\begin{equation}\label{eq:dis}
p(x) = \frac{A}{x^3}\exp\left[\frac{A}{2}\left(1-\frac{1}{x^2}\right)\right]\,,
\end{equation}
with
\begin{equation}
A=\frac{9\pi}{(Ze)^4}\frac{m^5}{{B'}^2(\epsilon'_i)^2\tau_0}\,.
\end{equation}

After decay we determine the distribution functions ${\rm d}N/{\rm d}\epsilon$ of secondary neutrinos and nucleons ($N$). The local neutrino emissivity can the be estimated as
\begin{equation}\label{eq:jnu}
j'_{\nu_\alpha}(\epsilon'_\nu) \simeq \sum_{\beta}P_{\alpha\beta}\int {\rm d}\epsilon'_p\left(\frac{1-e^{-\kappa\tau_{p\gamma}}}{\kappa}\right)\frac{{\rm d}N_{\nu_\beta}}{{\rm d} \epsilon'_\nu}(\epsilon'_p,\epsilon'_\nu)j'_p(\epsilon'_p)\,.
\end{equation}
where $P_{\alpha\beta}$ is the oscillation-averaged probability matrix of neutrino flavor transitions; see, {\it e.g.},~\cite{Bustamante:2019sdb}. The inelasticity $\kappa$ is here defined as
\begin{equation}\label{eq:x}
\kappa(\epsilon'_p) \equiv \int {\rm d}\epsilon''_N\frac{\epsilon'_p-\epsilon''_N}{\epsilon'_p}\frac{{\rm d}N_N}{{\rm d}\epsilon''_N}(\epsilon'_p,\epsilon''_N)\bigg/\int {\rm d}\epsilon''_N\frac{{\rm d}N_N}{{\rm d}\epsilon''_N}(\epsilon'_p,\epsilon''_N)\,.
\end{equation}
The definition (\ref{eq:jnu}) is consistent with energy relation of Eq.~(\ref{eq:Enu}).

% Don't change these lines
\bsp	% typesetting comment
\label{lastpage}
\end{document}